\documentclass[a4paper]{article}
\usepackage{multirow}
\usepackage{INTERSPEECH2021}
\usepackage{graphicx}
\usepackage{subfigure}
\title{CTA-RNN: Channel and Temporal-wise Attention RNN Leveraging Pre-trained ASR Embeddings for Speech Emotion Recognition}
\name{Chengxin Chen$^{1,2}$, Pengyuan Zhang$^{1,2}$}
\address{
  $^1$Key Laboratory of Speech Acoustics and Content Understanding, Institute of Acoustics, Chinese Academy of Sciences, China\\
  $^2$University of Chinese Academy of Sciences, China}
\email{\{chenchengxin, zhangpengyuan\}@hccl.ioa.ac.cn}

\begin{document}

\maketitle
\begin{abstract}
Previous research has looked into ways to improve speech emotion recognition (SER) by utilizing both acoustic and linguistic cues of speech.  However, the potential association between state-of-the-art ASR models and the SER task has yet to be investigated. In this paper, we propose a novel channel and temporal-wise attention RNN (CTA-RNN) architecture based on the intermediate representations of pre-trained ASR models.  Specifically, the embeddings of a large-scale pre-trained end-to-end ASR encoder contain both acoustic and linguistic information, as well as the ability to generalize to different speakers, making them well suited for downstream SER task. To further exploit the embeddings from different layers of the ASR encoder, we propose a novel CTA-RNN architecture to capture the emotional salient parts of embeddings in both the channel and temporal directions. We evaluate our approach on two popular benchmark datasets, IEMOCAP and MSP-IMPROV, using both within-corpus and cross-corpus settings. Experimental results show that our proposed method can achieve excellent performance in terms of accuracy and robustness.
\end{abstract}
\noindent\textbf{Index Terms}: speech emotion recognition, transfer learning, representation learning, information fusion

\section{Introduction}
Speech emotion recognition (SER), which aims to recognize the emotional characteristics of human speech and classify the utterance into one of a pre-defined set of emotion categories, has become a hot topic in the speech research field \cite{Schuller2018}. 
SER plays an important role in a variety of applications, including intelligent customer service and personal voice assistants.
To make human–machine interaction more natural and harmonic, it is vital for machines to capture the emotional changes of humans and respond in a proper way \cite{Picard1999}. 

In the era of deep learning, deep neural networks, such as convolutional neural networks (CNNs) \cite{Mao2014}, recurrent neural networks (RNNs) \cite{Lee2015}, and, more recently, self-attention mechanisms \cite{Mirsamadi2017},
%\cite{ZipingZhao2019}
%have become mainstream approaches for SER.
have been widely used for SER.
However, existing SER models are still far from human-level performance.
%However, it remains challenging to accurate recognize emotion from speech. 
%Human speech is a complicated signal that contains a wealth of linguistic and para-linguistic information. 
%There are two key factors that influence the perception of emotion. 
While low-level acoustic features, such as prosody, pitch, or tensity, can influence the preliminary perception of emotion, high-level semantics of speech aid in further confirming the speaker's real emotional state. 
%It is easy for human brains to process such complex information to perceive emotions comprehensively, but challenging for machines.
Human brains are capable of analyzing such complex information holistically to detect emotions, but machines have a hard time doing this.
To address this issue, researchers have proposed utilizing audio and text data simultaneously in recognizing emotions from speech. 
Specifically, the text modality could benefit from the large-scale pre-trained models (\emph{e.g.}, GloVe \cite{Glove} or BERT \cite{Devlin2019}) from the NLP area, resulting in more robust and accurate models. 
%The overview diagram of these methods is illustrated in Figure 1(a).
For example, Yoon \emph{et al.} \cite{Yoon2018} combined the high-level representations via concatenation and set up a baseline for bi-modal SER. 
Pan \emph{et al.} \cite{Pan2020} utilized a multi-modal attention sub-network to model inter-modality interactions before late fusion.
%More recently, Hu \emph{et al.} \cite{DongniHU2021} proposed a two-stage attention-based framework to model both intra- and inter-modality interactions.
Nevertheless, the majority of existing works assumed that the human transcriptions of speech were available from the start. 
In practice, the available transcriptions are derived from the output of an automatic speech recognition (ASR) module. 
%This matter may represent a bottleneck in a deployment pipeline due to computational complexity and potential errors in the transcription.
Hence, the computational complexity of the ASR module and potential errors in the transcriptions may represent a bottleneck in a deployment pipeline.

To solve this problem, several studies have explored how to incorporate linguistic information into the high-level representations of speech without explicit text input. 
%To solve this problem, several studies have explored ways to utilize linguistic information without explicit text input. 
%For example, Tzirakis \emph{et al.} \cite{Tzirakis2021} proposed to align the embedding spaces of Word2Vec and Speech2Vec models for linguistic feature extraction using only the speech signal.
For example, Cai \emph{et al.} \cite{XingyuCai2021} proposed a multi-task learning framework that performed ASR and SER at the same time to encode the linguistic information into the intermediate representations. 
Feng \emph{et al.} \cite{Feng2020} used the output of an ASR decoder as an alternative to the text features and combined it along with the conventional SER module. 
%Although these strategies demonstrated superiority on the SER benchmark datasets, the small amount of transcription made it difficult to develop a robust model.
%In this research, we propose an alternative to the frequently used spectrogram feature for robust SER: the intermediate representations of large-scale pre-trained end-to-end ASR models. 
However, most previous works only used rudimentary ASR models as an auxiliary part of their frameworks and failed to fully develop the capabilities of ASR models.
To date, there has been little research into the potential correlation between state-of-the-art ASR models and the SER task.
In this paper, we look into the viability of exploiting a large-scale pre-trained end-to-end ASR model, specifically, the Conformer model, for downstream SER task and compare the impact of ASR performance on SER.
Further, we propose a novel Channel and Temporal-wise Attention RNN (CTA-RNN) architecture to leverage the pattern discrepancy of embeddings from diverse blocks of the Conformer encoder.
Experimental results show that our proposed approach can achieve state-of-the-art performance on IEMOCAP and MSP-IMPROV. Moreover, the cross-corpus evaluations demonstrate a significant performance gain over the existing works.

\begin{figure*}[htbp]
  \centering
  \scalebox{0.8}
  {\includegraphics[width=\linewidth]{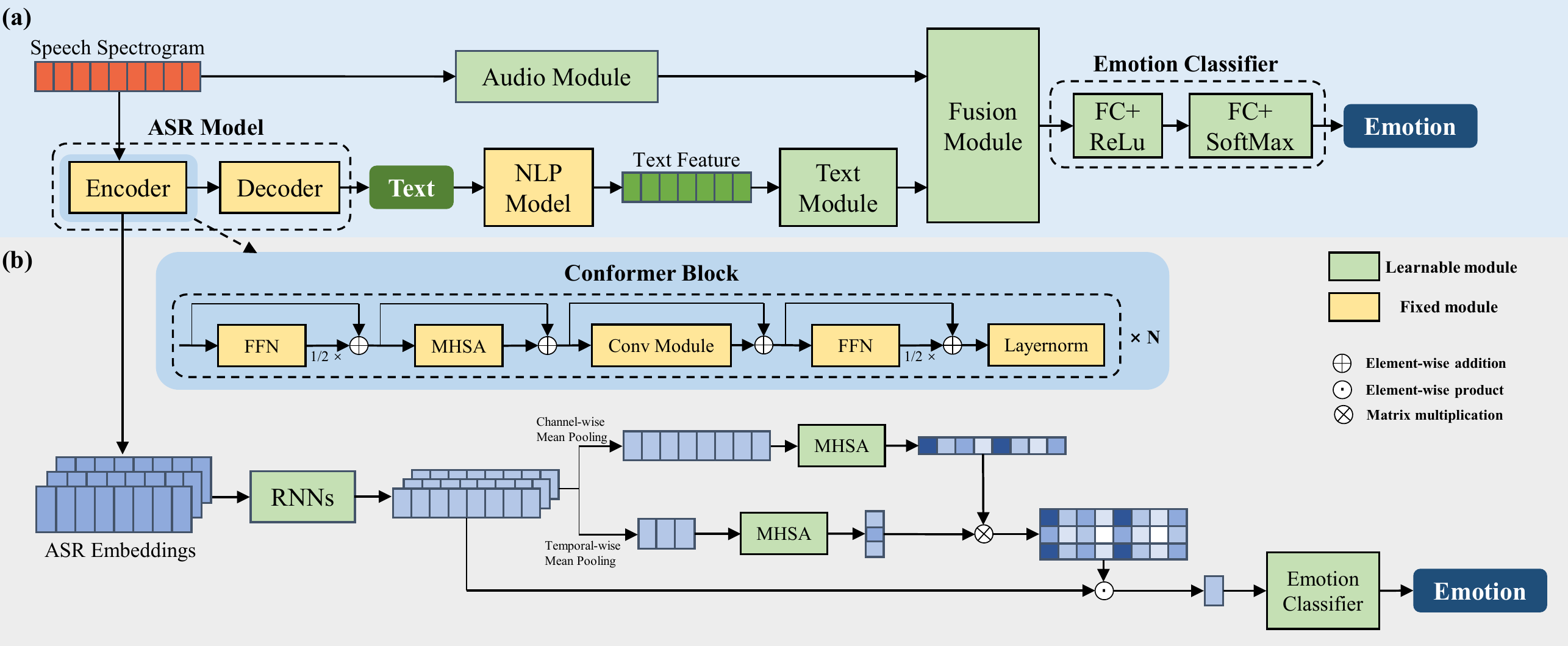}}
  \caption{The overview diagram of (a) the conventional pipeline for bi-modal SER and (b) our proposed methodology.}
  %{ The key difference is that we directly use the embeddings of the ASR encoder}
  \label{fig:overview}
\end{figure*}

\section{Related work}
%Mathematically, let $x_{i}$ be the input sequence to a Conformer block $i$, the output sequence $y_{i}$ of this block is:
%\begin{align}
%x_{i}'  & = x_{i} + \frac{1}{2} \text{FFN}(x_{i}), \\
%x_{i}'' & = x_{i}' + \text{MHSA} (x_{i}'), \\
%x_{i}''' & = x_{i}'' + \text{Conv} (x_{i}''), \\
%y_{i} & = \text{Layernorm} (x_{i}''' + \frac{1}{2} \text{FFN}(x_{i}''')),
%\end{align}
%In this way, the high-level representations encode both local context dependencies (captured by the convolution module) and the long context (captured by the self-attention module).
%In this way, the local context dependencies are captured by the convolution module, whilst the long context is captured by the self-attention module.
%In this paper, we use the pre-trained Conformer encoder as a front-end feature extractor
%We refer to the WeNet toolkit \cite{wenet} for the ASR training recipe of Conformer models. 
%In order to study the influence of ASR performance on downstream SER task, we compare the Conformer models pre-trained on a 960-hour subset of Librispeech \cite{librispeech} and a 10,000-hour subset of Gigaspeech \cite{gigaspeech}. 
%The pre-trained weights of the Conformer models can be downloaded at https://github.com/wenet-e2e/wenet/blob/main/docs/pretrained\_models.md.
\subsection{Conformer encoder}
Recently, the Conformer model \cite{Gulat2020} has been proposed for end-to-end ASR and achieved state-of-the-art performance. 
As shown in Figure 1, the Conformer encoder consists of N stacked Conformer blocks, where each block contains four modules arranged in a sandwich-like way: the Feed-Forward module (FFN), the Multi-Head Self-Attention module (MHSA), the Convolution module (Conv), and another Feed-Forward module. 
This architecture combines the advantages of Transformer models \cite{Zhang2020} and CNNs organically, being capable of capturing the long-term context and the local context dependencies.

\subsection{RNN with multi-head self-attention}
The architecture, RNN with multi-head self-attention (RNN-MHSA), has been widely used as a strong baseline in the task of SER \cite{Pan2020}\cite{Feng2020}\cite{DongniHU2021}. 
RNN is a prominent tool in modeling sequential data and works well at storing and accessing context information over a sequence. 
%In this work, we adopt a 2-layer bi-directional Gated Recurrent Unit (bi-GRU) networks. Each layer contains 128 hidden units followed by a dropout with probability of 0.5.
%Specifically, we feed the input spectrogram of an audio $\boldsymbol{A} = [a_1,a_2,...,a_n] \in \mathbb{R}^{n\times d_a}$
%a certain $\boldsymbol{E} \in \mathbb{R}^{m\times d_e}$ 
%into a 2-layer bi-directional Gated Recurrent Unit (bi-GRU), and concatenate the outputs of both forward and backward GRUs into $\boldsymbol{H} = [h_1,h_2,...,h_n]\in \mathbb{R}^{n\times d_h}$,
%The RNN module is applied on the input spectrogram of an audio $\boldsymbol{A} \in \mathbb{R}^{n\times d_a}$ to derive the 
%concatenated outputs of both forward and backward 
An RNN module is first applied on the spectrogram of speech to derive the state of hidden units $\boldsymbol{H} \in \mathbb{R}^{n\times d_h}$,
where $n$ is variable depending on the input length. 
%To capture the emotional salient parts of the sequence, the multi-head self-attention module computes
The self-attention module can capture the emotional salient parts of $\boldsymbol{H}$ in the temporal direction and output a fixed-length vector. 
Further, the multi-head structure can focus on different subspaces of the input representations at the same time. 
The output vectors from different heads are concatenated to construct a final vector before sent into the emotion classifier.

\section{Methodology}
\subsection{ASR Embeddings}
%We hypothesize that the intermediate representations of well pre-trained Conformer encoders contain both acoustic and linguistic information, and generalize well to unknown speakers. 
%due to the limitations of ASR models and datasets used in their experiments.
As demonstrated in Figure 1(a), a conventional bi-modal SER pipeline requires a well-trained ASR model to generate speech transcriptions first. 
%The high-level representations of audio and text are extracted before sent into a carefully designed fusion module.
The high-level representations of audio and text are then  extracted, before being sent into a carefully built fusion module.
%, and finally designs a suitable fusion method.
Differently, we propose using the ASR encoder's embeddings directly, as depicted in Figure 1(b).
The motivation stems from two aspects.
On the one hand, the state-of-the-art end-to-end ASR models \cite{e2easr1}\cite{e2easr2} integrate the acoustic model and language model into a single framework. 
Intuitively, the intermediate representations of a well-trained ASR model are capable of encoding both acoustic and latent linguistic information. 
On the other hand, the available datasets for ASR pre-training are large enough to ensure the robustness of ASR embeddings across different speakers, which may relieve the problem of present SER models' poor generality \cite{Poria2018}\cite{Latif2020}.
%Deep neural networks based SER models can suffer significantly when evaluated on the speakers out-of-domain.

Considering that fine-tuning the ASR models with the SER datasets can distort the pre-trained features and underperform them out-of-distribution, we fix the weights of a pre-trained ASR encoder and use it as a front-end feature extractor.
In this work, we focus on the encoder of the state-of-the-art Conformer models pre-trained on large-scale ASR datasets.
Given an input spectrogram $\boldsymbol{A} 
= [a_1,a_2,...,a_n]
\in \mathbb{R}^{n\times d_a}$, the ASR embeddings of the $i$-th Conformer block can be denoted as $\boldsymbol{E}^{(i)} =  [e_1,e_2,...,e_m] 
\in \mathbb{R}^{m\times d_e}$.
%, where $i\in \left \{ 1,2,\cdots ,N \right \}$.
%In the following parts, we will discuss how to utilize the embeddings of the conformer encoder for downstream SER. 
%In the following sections, we will discuss how to utilize $\boldsymbol{E}^{(i)}$ for downstream SER.
Hence, the problem can be formulated as how to map $\boldsymbol{E}^{(i)}$ into a fixed-length vector that is discriminative for SER, where $i\in \left \{ 1,2,\cdots ,N \right \}$.

%Since $\boldsymbol{E}^{(i)}$ is a type of sequential data similar to $\boldsymbol{A}$, 
%we can easily adapt RNN-MHSA to it.
%In the subsections that follow, we'll go through how to use the conformer encoder's embeddings for downstream SER. 

%The biggest disadvantage of this approach is the high cost of computation in terms of both time and memory.
%In order to take full advantage of the embeddings of a pre-trained Conformer encoder, we can stack the embeddings of different.
%Theoretically, as a neural network's layer depth increases, the pattern learnt becomes more abstract. 
%Intuitively, the shallow blocks of the Conformer tend to learn low-level patterns, while the deep blocks may learn more abstract patterns. 
%Deep learning allows computational models that are composed of multiple processing layers to learn representations of data with multiple levels of abstraction.\\

\subsection{CTA-RNN architecture}
%In order to fully utilize the pattern discrepancy of different Conformer blocks, we proposed an improved RNN model with both channel and temporal-wise attention (CTA-RNN). 
%We can directly use an $\boldsymbol{E}^{(i)}$ as the input to the RNN-MHSA model since $\boldsymbol{E}^{(i)}$ is a form of sequential data comparable to $\boldsymbol{A}$. When it comes to multi-channel case, i.e.
%The pattern encoded by the embeddings can get more abstract as the Conformer encoder goes deeper.
%When the Conformer encoder goes deeper, the pattern learned in different blocks' embeddings can become more abstract. 
%We can directly use $\boldsymbol{E}^{(i)} \in \mathbb{R}^{m\times d_e}$ as the input to an RNN-MHSA model since $\boldsymbol{E}^{(i)}$ is a form of sequential data. 
%The pattern learned by the embeddings can get more abstract as the Conformer encoder goes deeper, 
%Previous works have proved that fusing the embeddings from multiple layers of a pre-trained model can result in a more accurate and robust model, mainly because the pattern learned by different layers differs. 
Previous research has shown that combining the embeddings from multiple layers of a pre-trained model can result in a more accurate and robust downstream model, owing to the fact that the pattern learned by each layer differs \cite{Peters2018}\cite{chen2021wavlm}. 
A widely used strategy is to start with a trainable fully connected layer, and then utilize the weighted average of embeddings from different layers for the subsequent models. 
\iffalse
Mathematically, the combined embeddings can be calculated as:
\begin{align}
\widetilde{\boldsymbol{E}}   = \frac{\sum_{i  = 1}^{N} \alpha_{i} \boldsymbol{E}^{(i)}}{\sum_{i  = 1}^{N} \alpha_{i}}
\end{align}
where the learnt weights $\alpha_{i}$ can be thought of as the empirical importance of each layer. 
\fi
Hence, the learnt weights can be regarded as the empirical importance of each layer.
%However, the best weight distribution may vary for distinct utterances.
However, the best weight distribution for distinct utterances may differ.
%and even for different periods of utterance.
%The RNN-MHSA architecture was designed to focus on the emotional salient times of sequential data.

Inspired by the advantage of RNN-MHSA, we propose an RNN-based architecture for attending to the emotional salient parts of $\boldsymbol{E} \in \mathbb{R}^{N\times m\times d_e}$ in both the channel and temporal directions, dubbed CTA-RNN.
Given the embeddings $\boldsymbol{E} \in \mathbb{R}^{N\times m\times d_e}$, we first applied $N$ independent RNN modules to model the sequential data of each channel, 
%and we denote 
the combined state of hidden units is denoted as $\boldsymbol{H} \in \mathbb{R}^{N\times m\times d_h}$. 
An intuitive way to achieve both the channel and temporal-wise attention is to reshape the features into $\boldsymbol{H}' \in \mathbb{R}^{(N \cdot m)\times d_h}$ and applied a global self-attention.
However, the computation complexity will grow exponentially as $N$ increases.

To address this issue, we propose splitting the attention into two directions. We first applied channel and temporal-wise mean pooling to obtain two anchor features, denoted as $\boldsymbol{\widetilde{H}}_{c} \in \mathbb{R}^{m \times d_h}$ and $\boldsymbol{\widetilde{H}}_{t} \in \mathbb{R}^{N \times d_h}$, respectively. 
Afterwards, the MHSA modules are utilized to calculate the attention weights in the two directions.
Mathematically, the attention vector of $j$-th head can be computed as follows:
%Mathematically, the attention vector $\boldsymbol{\alpha}^j$ of the $j$-th head and the output $\boldsymbol{v}$ can be computed as follows:
\begin{align}
\boldsymbol{\alpha}_{t,j} & = \operatorname{SoftMax}\left( \frac{\boldsymbol{W}^{Q}_{t,j}(\boldsymbol{W}^{K}_{t,j} \boldsymbol{\widetilde{H}}_{t}^{T}) }{\sqrt{d_{\alpha } } } \right) \\
\boldsymbol{\alpha}_{c,j} & = \operatorname{SoftMax}\left( \frac{\boldsymbol{W}^{Q}_{c,j}(\boldsymbol{W}^{K}_{c,j} \boldsymbol{\widetilde{H}}_{c}^{T}) }{\sqrt{d_{\alpha } } } \right) 
\end{align}
where 
$\boldsymbol{W}^{Q}_{t,j}, \boldsymbol{W}^{Q}_{c,j} \in \mathbb{R}^{1\times d_{\alpha}}$ and 
$\boldsymbol{W}^{K}_{t,j}, \boldsymbol{W}^{K}_{c,j} \in \mathbb{R}^{d_{\alpha}\times d_{h}}$ are learnable  parameters, while
$\boldsymbol{\alpha}_{t,j}\in \mathbb{R}^{1 \times N}$ and 
$\boldsymbol{\alpha}_{c,j}\in \mathbb{R}^{1 \times m}$ are the scaled attention probability vectors.
Different from the conventional MHSA, we propose applying these attention vectors on the original $\boldsymbol{H} \in \mathbb{R}^{N\times m\times d_h}$. 
%$\boldsymbol{\alpha}_{t,j} \in \mathbb{R}^{d_{\alpha}}$ is the weighted sum of hidden states.
%Specifically, we first obtain the final attention matrices 
%$\boldsymbol{A}_{t}\in \mathbb{R}^{n_{\alpha} \times N}$ and 
%$\boldsymbol{A}_{c}\in \mathbb{R}^{n_{\alpha} \times m}$ by stacking the attention %vectors from $n_{\alpha}$ heads. 
%Specifically, the weighted sum of $\boldsymbol{H}$ of $j$-th head can be calculated as:
Specifically, the output $\boldsymbol{v}_j$ of the $j$-th head can be calculated as:
\begin{align}
\boldsymbol{A}_{j} & =\boldsymbol{\alpha}_{c,j} ^{T} \boldsymbol{\alpha}_{t,j} \\
\boldsymbol{v}_{j} &=  (\boldsymbol{W}^{V}_{j} \boldsymbol{H}^{T})\odot \boldsymbol{A}_{j} 
\end{align}
where $\boldsymbol{W}^{V}_{j} \in \mathbb{R}^{d_{\alpha} \times d_{h}}$ is trainable. $\boldsymbol{W}^{V}_{j}\boldsymbol{H}^{T} \in \mathbb{R}^{d_{\alpha} \times m\times N}$ represents the transformed “value feature”, while $\boldsymbol{A}_{j} \in \mathbb{R}^{m\times N}$ is the attention matrix.
We perform a element-wise multiplication (denoted as $\odot $) of $\boldsymbol{A}_{j}$ and the $d_{\alpha}$ sub-matrices of $\boldsymbol{W}^{V}_{j}\boldsymbol{H}^{T}$, respectively, and sum up each weighted matrices to derive the output $\boldsymbol{v}_{j} \in \mathbb{R}^{d_{\alpha}}$.
The final vector $\boldsymbol{v}\in \mathbb{R}^{n_{\alpha} \cdot d_{\alpha}}$ is obtained by concatenating the ouputs from $n_{\alpha}$ heads.

\section{Experiments}
\subsection{Datasets}
We use two popular benchmark datasets in English to evaluate the proposed approach.\\
\textbf{IEMOCAP:} The IEMOCAP dataset \cite{Busso2008} contains approximately 12 hours of audio-visual data from 10 actors. The dataset contains 5 sessions and each session is performed by one female and male actor in scripted and improvised scenarios, and the ground truth transcripts of speech are provided. To be consistent with the previous works, we merge the utterances labeled ‘excited’ into the ‘happy’ class, and the distribution of utterances used in the experiments is \{happy: 1636, sad: 1084, angry: 1103, neutral: 1708\}.\\
\textbf{MSP-IMPROV:} The MSP-IMPROV dataset \cite{Busso2017} contains 6 sessions of dyadic interactions between pairs of male-female actors. 
15 target sentences are used to collect the recordings. 
For each target sentence, 4 emotional scenarios were created to elicit happy, angry, sad and neutral responses. 
To make the emotion natural, two actors interacted with each other in improvised scenarios, and one person uttered the target sentences. 
The distribution of utterances used in the experiments is \{happy: 2644, sad: 885, angry: 792, neutral: 3477\}.
Compared with IEMOCAP, MSP-IMPROV contains more natural and spontaneous speech.
%and emotionally neutral transcriptions. 
Moreover, the emotion distribution of MSP-IMPROV is more unbalanced, making it a more challenging SER dataset.

\subsection{Experimental setup}
%\subsubsection{Data preprocessing}
\subsubsection{Implementation details}
We first resampled the speech signal at 16 \emph{kHz} and extracted the 80-dimensional Log Mel-scale Filter Bank (LMFB) with 25 \emph{ms} frame size  and 10 \emph{ms} frame shift.
The LMFB
%, denoted as $\boldsymbol{A} \in \mathbb{R}^{n\times 80}$, 
was used as the input to a pre-trained Conformer model to derive the embeddings of the encoder. 
In this work, we refer to the WeNet toolkit \cite{wenet} for the ASR training recipe of Conformer models. 
Specifically, the Conformer encoder consisted of 12 blocks and the embeddings were 512-dimensional.
In order to study the influence of ASR performance on downstream SER task, we compare the Conformer models pre-trained on a 960-hour subset of Librispeech \cite{librispeech} and a 10,000-hour subset of Gigaspeech \cite{gigaspeech}. 
For a fair comparison, we also performed ASRs on both SER datasets using the Conformer models to obtain speech transcriptions, which were further transformed into 1024-dimensional text features via a pre-trained BERT model.
%As presented in Table 1, the Conformer model pre-trained on Gigaspeech can significantly outperform the one pre-trained on Librispeech.
%The evaluation results of ASR performance are displayed in
Table 1 displays the evaluation results of ASR performance.
%The word error rate (WER) of the ASRs on IEMOCAP is 31.88\% (pre-trained on Librispeech) and 16.43\% (pre-trained on Gigaspeech), while WER on MSP-IMPROV is
%\subsection{Definition of experimental models}
%We denote the feature/embeddings used in the experiments as A (mel-spectrogram of Audio, 80-dim), T (BERT embeddings of Text, 1024-dim) and E (Embeddings of pre-trained Conformer based encoder, 512-dim), respectively.
%The uni-modal models consist of 
%The recurrent encoder based SER models are denoted as (*)-RE. For example, A-RE represents the model using A as input, while AT-RE represents the model using A and T as input.
%In this way, the target dataset is not used as prior knowledge, which is closer to the real situation.

%\subsubsection{Implementation details}
The hyper-parameters for training kept the same in all experiments.
The basic RNN-MHSA module consisted of a 2-layer Bi-GRU with 256 hidden units in each layer and a 8-head self-attention module with 64 nodes in each head.
The dropout of each Bi-GRU layer is 0.3. 
%By default, the model used for uni-modal input (\emph{e.g.}, $\boldsymbol{A}$ or $\boldsymbol{E}^{(i)}$) was RNN-MHSA, while the model used for bi-modal input ($\boldsymbol{A}+\boldsymbol{T}$) was LF-RNN-MHSA.
The batch size was set to 32, and the max training epochs were set to 50.
An Adam optimizer with an initial learning rate of $10^{-3}$ was applied to optimize the model parameters, and the learning rate was halved if the validation loss did not decrease for a consecutive 10 epochs.

\begin{table}[]
\caption{Evaluation results of ASRs on IEMOCAP and MSP-IMPROV. \textbf{ASR-PT}: Pre-trained ASR dataset. \textbf{WER}: Word error rate. \textbf{CER}: Character error rate.}
\label{tab:table1}
\centering
\scalebox{0.9}{
\begin{tabular}{ccccc}
\toprule
\multirow{2}{*}{\textbf{ASR-PT}} & \multicolumn{2}{c}{\textbf{IEMOCAP}} & \multicolumn{2}{c}{\textbf{MSP-IMPROV}} \\
\cmidrule(lr){2-3}\cmidrule(lr){4-5}
                        & WER(\%)          & CER(\%)          & WER(\%)            & CER(\%)           \\
\midrule                        
Librispeech             & $31.88$        & $22.28$        & $38.83$          & $27.24$         \\
Gigaspeech              & $\mathbf{16.43}$        & $\mathbf{12.30}$        & $\mathbf{18.68}$          & $\mathbf{13.85}$         \\
\bottomrule 
\end{tabular}
}
\end{table}

\subsubsection{Baseline models}
We denote the pre-processed sequential features used in the following experiments as 
$\boldsymbol{A} \in \mathbb{R}^{n \times 80}$, 
$\boldsymbol{T} \in \mathbb{R}^{w \times 1024}$, and
$\boldsymbol{E^{(i)}} \in \mathbb{R}^{m \times 512}$.\\
%, where $i\in \left \{ 2, 4, 6, 8, 10, 12 \right \}$.\\
%\subsection{Methods for comparison}
%Although RNN with multi-head self-attention has been proved to be effective for time-varying sequential input into a fixed-length vector for the following classifier.
%Although RNN with multi-head self-attention has been proven to be effective for converting single-channel time sequential input into a fixed-length vector,
%RNN with multi-head self-attention can be easily adapted to the sequential data $\boldsymbol{E}^{(i)} \in \mathbb{R}^{m\times d_e}$.
%, where $i\in \left \{ 1,2,\cdots ,N \right \}$. 
%However, further research is needed to determine which Conformer block's output is best for downstream SER. 
%An easier way is to fuse the overall output embeddings E
\textbf{RNN:} 
It represents the baseline approach that directly uses $\boldsymbol{A}$, $\boldsymbol{T}$ or $\boldsymbol{E}^{(i)}$ as the input to an RNN-MHSA model. \\
%We directly use an $\boldsymbol{E}^{(i)}$ as the input to the RNN-MHSA model since $\boldsymbol{E}^{(i)}$ is a form of sequential data comparable to $\boldsymbol{A}$. \\
%Intuitively, the bottom blocks tend to encode more low-level features, while the top blocks may learn more abstract patterns.
%Intuitively, as the conformer encoder goes deeper, the output embeddings tend to encode more abstract patterns.
%Intuitively, the pattern encoded in $\boldsymbol{E}^{(i)}$ may change from low-level acoustic features to high-level linguistic features as the block number $i$ increases.
%However, further investigation is needed to determine which $\boldsymbol{E}^{(i)}$ is the best input for SER. \\
%This approach also proves to be effective in conventional bi-modal SER with the ability to handle asynchronous data streams. \\
%Another feasible approach is deploying $N$ independent RNN-MHSA models to convert each $\boldsymbol{E}^{(i)}$ into a fixed-length vector $\boldsymbol{v}^{(i)}$ and concatenate them finally, dubbed late fusion (LF). 
%This approach proves to be effective in conventional bi-modal SER with the ability to handle asynchronous data streams. \\
\textbf{WF-RNN:} 
A trainable fully connected layer is fist applied to compute the weighted average of embeddings from different blocks, dubbed weighted fusion (WF). 
The result $\widetilde{\boldsymbol{E}}$ will be sent into an RNN-MHSA model.\\
%used as the input to \\
\textbf{EF-RNN:}
The ASR embeddings from different blocks are first stacked in the dimensional direction to construct $\boldsymbol{E}'\in \mathbb{R}^{m \times 6144}$, dubbed early fusion (EF). 
The result $\boldsymbol{E}'$ will be sent into an RNN-MHSA model.
%In order to fully exploit the ASR embeddings, we first stack them to construct $\boldsymbol{E} \in \mathbb{R}^{m\times (N \cdot d_e)}$, dubbed early fusion (EF), and take $\boldsymbol{E}$ as the input to the RNN-MHSA model.
%This method is simple, but the high-dimensional feature set may easily suffer from the problem of data sparseness.
\\
%Inspired by the method used in \cite{Peters2018} for NLP, 
%Different from EF-RNN-MHSA, we can apply a trainable fully connected layer to compute the weighted average of embeddings from different blocks, dubbed weighted fusion (WF). 
%The result $\widetilde{\boldsymbol{E}} \in \mathbb{R}^{m\times d_e}$ will be used as the input to the RNN-MHSA model, and the weight of different blocks can be regardrd as the importance of embeddings.
\textbf{LF-RNN:} 
Several independent RNN-MHSA models are deployed to convert each sequential data into a fixed-length vector, respectively
%$\boldsymbol{v}^{(i)}$ 
%and concatenate them finally
, dubbed late fusion (LF). 
The vectors are concatenated before being sent into an RNN-MHSA model.
This approach also proves to be effective in conventional bi-modal SER with the ability to handle time-asynchronous data streams. 

\subsubsection{Evaluation settings}
To evaluate the generalization of the models, we performed both within-corpus (denoted as IEM and MSP) and cross-corpus (denoted as IEM2MSP and MSP2IEM) experiments.
For within-corpus SER, we applied cross-validation (CV) to avoid the impact of limited data. 
Take IEMOCAP for example, a 10-fold leave-one-speaker-out CV was conducted, \emph{i.e.}, 4 sessions from 8 speakers were used for training, while utterances from the two speakers in the left session were used for validating and testing, respectively. 
Similarly, a 12-fold leave-one-speaker-out CV was conducted on MSP-IMPROV. 
While for cross-corpus SER, we directly evaluated the best models trained from the within-corpus experiments on the other dataset. 
For instance, we used all the data from MSP-IMRPOV to evaluate the 10 best models trained on IEMOCAP and averaged the results, and vice versa. 
Since the two datasets are unbalanced, we utilize the officially recommended metric,  named unweighted average recall (UAR). 
UAR is insensitive to the impact of class imbalance and averages the sum of each class recall by the number of classes. 
%To enhance the reliability of the results, we repeated the experiments with five different random seeds for initialization and reported the mean and standard deviation of UAR.

\subsection{Experimental results and discussion}
\subsubsection{Feasibility analysis of ASR embeddings}

\begin{figure}[htbp]
  \centering
  \scalebox{0.99}{
  \includegraphics[width=\linewidth]{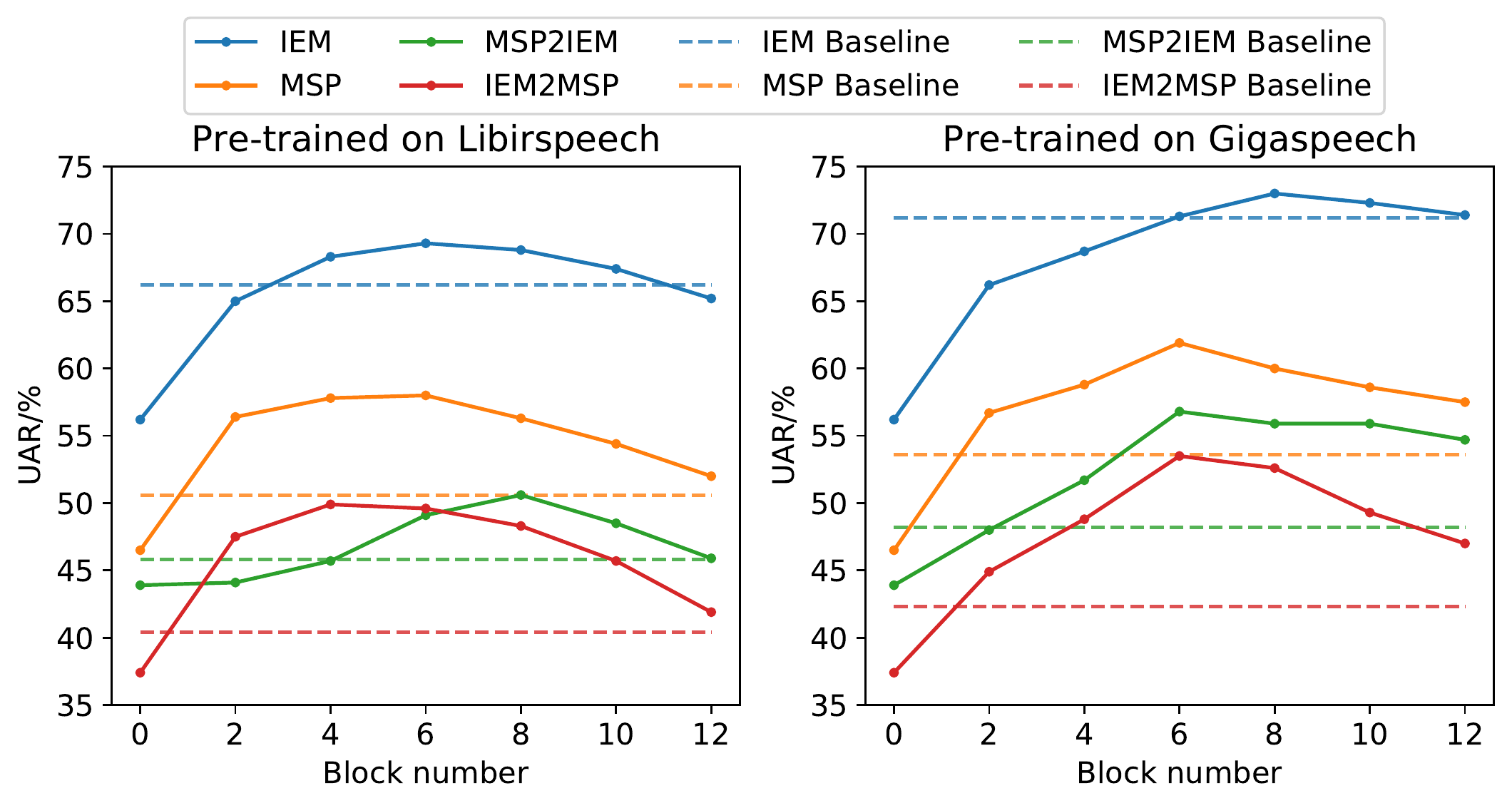}}
  \caption{Within-corpus and cross-corpus experimental results using the Conformer models pre-trained on Librispeech (left) and Gigaspeech (right).
  Dashed lines represent the baseline models for bi-modal SER.
  Solid lines represent the models based on ASR embeddings from different blocks, where block 0 represents the original LMFB. }
  %\textbf{IEM}: trained and tested on IEMOCAP.
  %\textbf{IEM2MSP}: trained on IEMOCAP and tested on MSP-IMPROV.}
  %{ The key difference is that we directly use the embeddings of the ASR encoder}
  \label{fig:result}
\end{figure}

%Compared with IEMOCAP, MSP-IMPROV contains more natural and spontaneous speech.
%Moreover the emotion distribution of MSP-IMPROV is more unbalanced.
%We first compare the effects of the ASR systems with different performance on downstream SER tasks. The WER of 
%We first compare the baseline models that using different input features, and the results are presented in the Table 2. 
%We first compare the performance of models using different features.
We first compare the performance of three types of models using different features as input: (1) $\boldsymbol{E^{(i)}}$ models: RNN models using $\boldsymbol{E^{(i)}}$ as input;
(2) $\boldsymbol{A}$ models: RNN models using $\boldsymbol{A}$ as input;
(3) $\boldsymbol{A}+\boldsymbol{T}$ models: LF-RNN models using both $\boldsymbol{A}$ and $\boldsymbol{T}$ as inputs.
In this subsection, we will focus on the $\boldsymbol{E^{(i)}}$ where $i\in \left \{ 2, 4, 6, 8, 10, 12 \right \}$.
As demonstrated in Figure 2, the $\boldsymbol{E^{(i)}}$ models (plotted in solid lines) can consistently outperform the $\boldsymbol{A}$ models (plotted at block 0).
When the block number $i$ increases from 2 to 12, the performance of $\boldsymbol{E^{(i)}}$ models show an upward trend followed by a downward trend.
In most cases, the $\boldsymbol{E^{(i)}}$ models can reach an optimal performance when $i$ equals 6.
%using the embeddings from the middle blocks.
Moreover, the $\boldsymbol{E^{(i)}}$ models can outperform the $\boldsymbol{A}+\boldsymbol{T}$ models (plotted in dashed lines as baseline) under the same pre-trained Conformer models.
The performance gain is even more significant in the cross-corpus settings, demonstrating the advantage of ASR embeddings in terms of robustness.
Comparing the two pre-trained Conformer models, a higher UAR for downstream SER can be achieved with a lower WER of ASR, both for the $\boldsymbol{A}+\boldsymbol{T}$ models and the $\boldsymbol{E^{(i)}}$ models.
Further experiments will only consider Gigaspeech based ASR-embeddings.

\begin{table}[htbp]
%\caption{Performance comparison of $\boldsymbol{E^{(i)}}$ based RNN-MHSA models and $\boldsymbol{E^{(i)}} + \boldsymbol{A}/\boldsymbol{T}$ based LF-RNN-MHSA models.}
\caption{Performance comparison of $\boldsymbol{E^{(i)}}$, $\boldsymbol{E^{(i)}} + \boldsymbol{A}$ and $\boldsymbol{E^{(i)}} + \boldsymbol{T}$ models, where $i\in \left \{ 2, 6, 12 \right \}$.} 
\label{tab:table3}
\centering
\scalebox{0.8}{
\begin{tabular}{lcccc}
\toprule
\multirow{2}{*}{\textbf{Input}} & \multicolumn{2}{c}{\textbf{Within-corpus UAR(\%)}} & \multicolumn{2}{c}{\textbf{Cross-corpus UAR(\%)}} \\
\cmidrule(lr){2-3}\cmidrule(lr){4-5}
                        & IEM             & MSP             & MSP2IEM             & IEM2MSP            \\
\midrule                       
$E^{(2)}$   & 66.2           & 56.7          & 48.0           & 44.9          \\
%                       & 4                      & $68.7\pm1.2$           & $58.8\pm1.5$           & $51.7\pm0.6$           & $48.8\pm1.1$          \\
$E^{(6)}$                       & 71.3           & \textbf{61.9}  & \textbf{56.8}  & \textbf{53.5} \\
%                       & 8                      & $\mathbf{73.0\pm0.3}$  & $60.0\pm0.4$           & $55.9\pm0.6$           & $52.6\pm0.4$          \\
%                       & 10                     & $72.3\pm1.2$           & $58.6\pm0.6$           & $55.9\pm0.6$           & $49.3\pm0.5$          \\
$E^{(12)}$                       & \textbf{71.4}           & 57.5          & 54.7           & 47.0          \\
\midrule                       
$A+E^{(2)}$                       & 65.4           & 56.4           & 47.8           & 43.6          \\
$A+E^{(6)}$                       & 71.4          & \textbf{61.6}  & \textbf{57.3}  & \textbf{53.7} \\
$A+E^{(12)}$                        & \textbf{72.3}  & 58.0         & 56.4           & 47.7          \\
\midrule                       
$T+E^{(2)}$                      & 74.9           & 61.1           & 51.8           & 48.0          \\
$T+E^{(6)}$                       & \textbf{76.5}  & \textbf{63.0}  & \textbf{59.2}  & \textbf{55.1} \\
$T+E^{(12)}$                       & 72.4           & 57.9           & 52.6           & 46.4         \\
\bottomrule                       
\end{tabular}
}
\end{table}

To further investigate the pattern discrepancy of different blocks, we performed
%We consider the representative ASR embeddings from 
experiments on $\boldsymbol{A}+\boldsymbol{E^{(i)}}$ and $\boldsymbol{T}+\boldsymbol{E^{(i)}}$ models, constructed in a similar way as $\boldsymbol{A}+\boldsymbol{T}$ models.
Three representative $\boldsymbol{E^{(i)}}$ were selected for comparison.
The experimental results are reported in Table 2. 
Compared with the $\boldsymbol{E^{(i)}}$ baseline models, 
%incorporating $\boldsymbol{A}$ features cannot bring obvious performance gain.
%the $\boldsymbol{T}+\boldsymbol{E^{(i)}}$ models
integrating $\boldsymbol{T}$ features can further boost the performance, whereas incorporating $\boldsymbol{A}$ features can achieve little performance gain, indicating that the majority of acoustic emotional cues have been preserved in $\boldsymbol{E^{(i)}}$.
Another finding is that the $\boldsymbol{T}+\boldsymbol{E^{(2)}}$ models can achieve better results than the $\boldsymbol{T}+\boldsymbol{E^{(12)}}$ models, which is in opposite to the cases of $\boldsymbol{E^{(i)}}$ models and $\boldsymbol{A}+\boldsymbol{E^{(i)}}$ models.
%although both $\boldsymbol{A}+\boldsymbol{E^{(i)}}$ and $\boldsymbol{T}+\boldsymbol{E^{(i)}}$ models can achieve the best performance when $i$ equals 6, 
%the $\boldsymbol{A}+\boldsymbol{E^{(12)}}$ models can outperform the $\boldsymbol{A}+\boldsymbol{E^{(2)}}$ models, which is consistent with the case of $\boldsymbol{E^{(i)}}$ models. 
%By contrast, the $\boldsymbol{T}+\boldsymbol{E^{(2)}}$ models can achieve better results than the $\boldsymbol{T}+\boldsymbol{E^{(12)}}$ models.
%One probable explanation is that the ASR embeddings at the bottom of encoder contain more low-level acoustic features, whereas those at the top of encoder can learn more abstract linguistic information.
One probable explanation is that the ASR embeddings at the top of encoder can learn more abstract linguistic information than those at the bottom of encoder, whereas low-level acoustic patterns can be partially lost at the top of encoder.
%One probable explanation is that the ASR embeddings at the bottom of  encoder lack high-level linguistic information, whereas low-level acoustic patterns can be partially lost at the top of encoder.
As a result, the ASR embeddings at the middle blocks can be regarded as a compromise between acoustic and linguistic information, providing a more accurate and robust representation for downstream SER.

\subsubsection{Fusion of ASR embeddings}
%Although the superiority of large-scale pre-trained ASR embeddings have been validated above, we further 
In Table 3, we present the experimental results of different fusion approaches to take full advantage of $\boldsymbol{E^{(i)}}$, where $i\in \left \{ 1,2,\cdots,12 \right \}$.
%embeddings from different blocks.
Comparing the baseline models, we find that late fusion (LF) performs better in within-corpus experiments, whereas early fusion (EF) performs better in cross-corpus evaluation. 
%our proposed CTA-RNN models demonstrate the best performance both in within-corpus and cross-corpus experiments.
Overall, our proposed CTA-RNN models demonstrate the best performance in all experimental settings. 
We also perform experiments that applied CTA directly on the embeddings without RNN modules.
%This modification leads to a slight performance decay. 
A slight performance degradation can be observed as a result of this change.
In return, 
%the total complexity of calculation and training time can be significantly lowered.
the overall computation complexity and training time can be lowered by a large margin.
Compared with existing state-of-the-art approaches, our proposed CTA-RNN architecture can significantly improve the performance of SER in both within-corpus and cross-corpus experiments.
Different from previous works \cite{Latif2020}\cite{Abdel2018}\cite{Ahn2021}, 
%the cross-corpus experiments performed in this paper did NOT access to the unlabeled target datasets in advance.
the unlabeled target datasets were not available in advance for the cross-corpus experiments described in this paper.
The excellent robustness of our approach mainly benefits from the large-scale ASR datasets for pre-training.

\begin{table}[]
\caption{Performance comparison of our proposed CTA-RNN and the baseline models. State-of-the-art results of previous approaches are also presented for reference.}
\label{tab:table4}
\centering
\scalebox{0.8}{
\begin{tabular}{ccccc}
\toprule
\multirow{2}{*}{\textbf{Method}}  & \multicolumn{2}{c}{\textbf{Within-corpus UAR(\%)}} & \multicolumn{2}{c}{\textbf{Cross-corpus UAR(\%)}} \\
\cmidrule(lr){2-3}\cmidrule(lr){4-5}
                       & IEM             & MSP             & MSP2IEM             & IEM2MSP            \\
                       
%\textbf{Method}  & \textbf{IEM}      & \textbf{MSP}      & \textbf{MSP2IEM}  & \textbf{IEM2MSP}  \\
\midrule 
Feng et al. \cite{Feng2020} & 69.7 & - & - & - \\
Lu et al. \cite{Lu2020} & 72.6 & - & - & - \\
Latif et al. \cite{latif2020multi} & 66.7 & 60.2 & - & - \\
Yoon et al. \cite{Yoon2018} & 73.1 & 54.0 & 48.5 & 41.8 \\
Abdel et al. \cite{Abdel2018}  & - & - & 53.1 & 42.9 \\
Ahn et al. \cite{Ahn2021} & - & 46.0 & - & 45.8 \\
Latif et al.\cite{Latif2020} & 64.1 & 56.2 & - & 46.8 \\
\midrule 
WF-RNN  & 72.8 & 61.1 & 56.5 & 52.1 \\
EF-RNN  & 74.4 & 61.9 & 59.1 & 52.1 \\
LF-RNN  & 75.3 & 62.8 & 58.2 & 50.7 \\
\midrule 
CTA w/o RNN & 75.6 & 62.4 & 58.9 & 52.9 \\
CTA-RNN & \textbf{76.1} & \textbf{63.5} & \textbf{59.4} & \textbf{54.3} \\
\bottomrule   
\end{tabular}
}
\end{table}

\iffalse
\begin{table}[htbp]
\caption{Table4}
\label{tab:table4}
\centering
\scalebox{0.8}{
\begin{tabular}{ccccc}
\toprule
\multirow{2}{*}{\textbf{Model}}  & \multicolumn{2}{c}{\textbf{Within-corpus UAR(\%)}} & \multicolumn{2}{c}{\textbf{Cross-corpus UAR(\%)}} \\
\cmidrule(lr){2-3}\cmidrule(lr){4-5}
                       & IEM             & MSP             & MSP2IEM             & IEM2MSP            \\
                       
%\textbf{Method}  & \textbf{IEM}      & \textbf{MSP}      & \textbf{MSP2IEM}  & \textbf{IEM2MSP}  \\
\midrule 
WF-RNN  & $72.8\pm1.0$ & $61.1\pm1.1$ & $56.5\pm1.8$ & $52.1\pm1.4$ \\
EF-RNN  & $74.4\pm0.8$ & $61.9\pm0.9$ & $59.1\pm0.9$ & $52.1\pm0.5$ \\
LF-RNN  & $75.3\pm0.8$ & $62.8\pm0.6$ & $58.2\pm0.6$ & $50.7\pm0.7$ \\
\midrule 
CTA-RNN & $\mathbf{76.2\pm0.6}$ & $\mathbf{63.5\pm0.7}$ & $\mathbf{59.8\pm0.5}$ & $\mathbf{54.4\pm0.6}$ \\
\bottomrule   
\end{tabular}
}
\end{table}
\fi

\section{Conclusions}
In this paper, we investigate the viability of leveraging the embeddings of a large-scale pre-trained Conformer encoder for downstream SER.
%Experimental results indicate that embeddings from the bottom of encoder contain more acoustic features, whereas the embeddings from the top of encoder contain more linguistic information.
Empirically, the embeddings from the middle blocks of a well pre-trained ASR encoder may contain adequate acoustic and linguistic information, making them more suited for SER.
Additionally, improving the ASR performance of pre-trained models can benefit for downstream SER.
%Empirically, the ASR embeddings from the middle blocks of encoder are the best choice.
Further, we propose a CTA-RNN architecture to effectively fuse the ASR embeddings from different blocks. 
The experimental results show that our proposed methodology can achieve state-of-the-art performance on IEMOCAP and MSP-IMPROV without explicit text as input.
Moreover, the cross-corpus evaluations demonstrate the robustness of large-scale pre-trained ASR embeddings across different speakers and domains.
In future work, we intend to explore the correlation of emotional speech in several languages and expand on our work to multi-lingual SER.

\bibliographystyle{IEEEtran}
\footnotesize
%\tiny
\bibliography{mybib}

% \begin{thebibliography}{9}
% \bibitem[1]{Davis80-COP}
%   S.\ B.\ Davis and P.\ Mermelstein,
%   ``Comparison of parametric representation for monosyllabic word recognition in continuously spoken sentences,''
%   \textit{IEEE Transactions on Acoustics, Speech and Signal Processing}, vol.~28, no.~4, pp.~357--366, 1980.
% \bibitem[2]{Rabiner89-ATO}
%   L.\ R.\ Rabiner,
%   ``A tutorial on hidden Markov models and selected applications in speech recognition,''
%   \textit{Proceedings of the IEEE}, vol.~77, no.~2, pp.~257-286, 1989.
% \bibitem[3]{Hastie09-TEO}
%   T.\ Hastie, R.\ Tibshirani, and J.\ Friedman,
%   \textit{The Elements of Statistical Learning -- Data Mining, Inference, and Prediction}.
%   New York: Springer, 2009.
% \bibitem[4]{YourName17-XXX}
%   F.\ Lastname1, F.\ Lastname2, and F.\ Lastname3,
%   ``Title of your INTERSPEECH 2021 publication,''
%   in \textit{Interspeech 2021 -- 20\textsuperscript{th} Annual Conference of the International Speech Communication Association, September 15-19, Graz, Austria, Proceedings, Proceedings}, 2020, pp.~100--104.
% \end{thebibliography}

\end{document}